\documentclass[aps,prf,amsmath,amssymb,longbibliography,superscriptaddress]{revtex4-1}

\usepackage{amsmath}
\usepackage{amssymb}
\usepackage{wasysym}
\usepackage{pifont}
\usepackage{graphicx,overpic,xcolor}
\usepackage{mathrsfs}
\newcommand{\ve}[1]{\ensuremath{\mbox{\boldmath$#1$}}}
\newcommand{\ma}[1]{\ensuremath{\mathbb{#1}}}

\newcommand{\T}{{\small \sf T}}
\newcommand{\ddt}{\tfrac{{\rm d}}{{\rm d}t}}

\newcommand{\eqnlab}[1]{\label{eq:#1}}

\newcommand{\eqnref}[1]{{(\ref{#1})}}

\newcommand{\Eqnref}[1]{{Eq.~(\ref{#1})}}

\newcommand{\marker}[1]{\protect\includegraphics[width=2mm,clip]{markBW#1.pdf}}

\DeclareMathOperator{\ku}{Ku}
\DeclareMathOperator{\st}{St}

\begin{document}

\title{Passive directors in turbulence}
\author{L. Zhao\textsuperscript{$\ast$}}
\address{\hspace*{-2mm}{\textsuperscript{\colorbox{white}{\phantom{X}}}}\textsuperscript{$\ast$}These authors contributed equally to this work. \\
\mbox{\mbox{}$^1$Department} of Engineering Mechanics, Tsinghua University, 100084 Beijing, China}
\address{Department of Energy and Process Engineering, NTNU, NO-7491 Trondheim, Norway}
\author{K. Gustavsson\textsuperscript{$\ast$}}
\address{Department of Physics, Gothenburg University, SE-41296 Gothenburg, Sweden}
\author{R. Ni}
\address{Department of Mechanical Engineering, Johns Hopkins University, Baltimore, Maryland 21218, USA }
\author{S. Kramel} 
\address{Department of Physics, Wesleyan University, Middletown, Connecticut 06459, USA}
\author{G. A. Voth}
\address{Department of Physics, Wesleyan University, Middletown, Connecticut 06459, USA}
\author{H. I. Andersson}
\address{Department of Energy and Process Engineering, NTNU, NO-7491 Trondheim, Norway}
\author{B. Mehlig}
\address{Department of Physics, Gothenburg University, SE-41296 Gothenburg, Sweden}

\begin{abstract}
In experiments and numerical simulations we measured angles between the symmetry axes of small spheroids advected in 
turbulence ({\em passive directors}). Since turbulent strains tend to align nearby spheroids, one might think that their relative angles are quite small. We show that this intuition fails in general because angles between the symmetry axes of nearby particles are anomalously large. We identify two mechanisms that cause this phenomenon. First, the dynamics evolves to a fractal attractor despite the fact that the fluid velocity is spatially smooth at small scales. Second, 
this fractal forms steps akin to scar lines observed in the director patterns for random or chaotic two-dimensional maps.
\end{abstract}

\maketitle

\section{Introduction}
Suspensions of small particles in turbulence determine the physics and chemistry of many natural processes.  The analysis of the underlying  highly non-linear and multi-scale dynamics poses formidable challenges, because any description of the problem must refer to the turbulence that the particles experience as they move through the fluid. Experiments resolving the particle dynamics have only recently become possible, and direct numerical simulations (DNS) of such systems are still immensely difficult.  Recently there has been substantial progress in understanding the dynamics of {\em spherical} particles in turbulence by  means of statistical models \cite{Fal01,Gus16a}.

Yet most solid particles we encounter in Nature and engineering are not spherical, such as ice crystals in turbulent clouds \cite{Pru78}, plankton in the turbulent ocean \cite{Gua12,Dur13,Ber16,Sen17}, and turbulent fibre flows in industrial processing \cite{Lun11}. Therefore it is necessary to understand how non-spherical particles translate and rotate in turbulence \cite{Voth16}. For very small particles, inertial effects are negligible \cite{Can16}, and the disturbance caused by the particles can be treated in the Stokes approximation \cite{Jef22,Toschi,Voth16}. To understand the turbulent angular dynamics of non-spherical particles in this limit is a question of great current interest \cite{Gir90,Pum11,Par12,Che13,Gus14b,Ni14,Voth14,Voth16}. But even the angular dynamics of a single small rod in turbulence is quite intricate: rods tend to align with the local vorticity of the flow \cite{Tabor94,Pum11,Ni14}. Vorticity in turn aligns with the second eigenvector of the turbulent strain-rate matrix \cite{xu2011}, and picks up that turbulence breaks time-reversal invariance \cite{jucha2014,pumir2016}. Polymers tend to align with the main stretching direction of the
flow \cite{Fou03}.

Very little is known about how non-spherical particles orient relative to each other in a turbulent flow, 
even in the limit of very small particles whose centre-of-mass position $\ve x$ simply follows the flow  ({\em passive} particles).
How do  nearby non-spherical particles align with each other? The simplest case is that of axisymmetric particles with fore-aft symmetry. The symmetry axes     $\ve n$    of an ensemble of such particles in turbulence form a spatial field, $\ve n(\ve x,t)$. 
At every point in space and time $\ve n$ is normalised to unity.
Our goal is to determine the geometrical properties of this field. For fore-aft symmetric particles the problem is invariant under $\ve n \to -\ve n$, so that $\ve n(\ve x,t)$ is in effect a field of {\em directors}.
It is plausible that turbulent strains align the particles as they approach, and in this case one
expects the {\em passive} director field $\ve n(\ve x,t)$ to be  a smooth function of particle position. But in this paper we show that the 
spatial passive-director patterns of non-spherical particles in turbulence are not smooth in general, not even at the smallest scales where the turbulent fluid velocities are smooth functions of position.  We show that  angles between the symmetry axes of nearby particles are anomalously large. Our results of DNS and statistical-model simulations allow us to conclude, first,  that the attractor determining the director  patterns is fractal, in general. Second,  the steady-state distribution of angles between nearby particles has power-law tails. We derive a theory based on diffusion approximations that can at least qualitatively explain these observations, and we relate the power-law tails to 
 {\em steps} of different widths that occur  in the  director patterns.

\section{Formulation of the problem}
The centre-of-mass positions $\ve x$ of small particles simply follow the flow $\ve u(\ve x,t)$,
\begin{equation}
\label{eq:adv}
\tfrac{{\rm d}}{{\rm d}t}\ve x
= \ve u(\ve x,t)\,,
\end{equation}
if their spatial diffusion is neglected.
In this case one says that the centre-of-mass is {\em advected}  by the flow \cite{Fal01}. The directors of small axisymmetric particles with fore-aft symmetry obey Jeffery's equation \cite{Jef22}
\begin{equation}
\label{eq:jeffery}
 \tfrac{{\rm d}}{{\rm d}t}\ve n =
\ma O \ve n + \Lambda \ma S\ve n-\Lambda (\ve n\cdot \ma S\ve n)\ve n\,.
\end{equation}
Here $\ma O$ is the anti-symmetric part of the matrix $\ma A$ of fluid-velocity gradients at the particle position, $\ma S$ is its symmetric part.
Inertial effects \cite{Can16} and angular diffusion \cite{hinch1972} are neglected in Eq.~(\ref{eq:jeffery}).
The parameter $\Lambda$ parameterises particle shape \cite{Bretherton:1962,Jef22}: $\Lambda\!=\!0$ for spheres, $\Lambda\!=\!-1$ for thin disks, and $\Lambda\!=\!1$ for slender rods.
 For spheroids
  \begin{equation}
\label{eq:Lambda}
\Lambda = (\kappa^2-1)/(\kappa^2+1)\,,
\end{equation}  where  $\kappa$ is the aspect ratio of the spheroidal particle  \cite{Jef22}.
In the following we consider prolate particles with $\Lambda\geq 0$.
We use two different kinds of simulations to analyse the director patterns $\ve n(\ve x,t)$. First, we employ DNS of a turbulent channel flow to obtain a turbulent velocity field, and integrate
Eqs.~\eqnref{eq:adv} and \eqnref{eq:jeffery} numerically to determine
the director patterns. Second, we perform simulations of Eqs.~\eqnref{eq:adv} and \eqnref{eq:jeffery} using a statistical model, representing the small-scale turbulent velocities as a random Gaussian incompressible, homogeneous, and isotropic field with correlation length $\eta$, correlation time $\tau$, and rms speed $u_0$. The theory employs diffusion
approximations that are valid in the limit of small {\em Kubo number}
$\ku\equiv u_0\tau/\eta$ and that have yielded important insights into the dynamics of small spherical particles in turbulence \cite{Gus16a}, just like Kraichnan's diffusion model  for passive-scalar advection \cite{Fal01}. 
Our experiments measure the angles between symmetry axes of fibres advected in a turbulent flow between oscillating grids.

To characterise the director patterns $\ve n(\ve x,t)$ we measure the statistics of $\delta\ve n(\ve x,t) \equiv \ve n(\ve x+\ve R,t)\pm \ve n(\ve x,t)$ at small distances $R \equiv |\ve R|$.  The configurations $\pm \ve n(\ve x,t)$ correspond to identical physical situations, and we choose the sign so that $|\delta\ve n|$ is minimal.  We define the \lq angular structure functions\rq{}
\begin{equation}
\label{eq:spr}
S_p(R) \equiv \langle |\delta\ve n|^p\rangle_R\,.
\end{equation}
Here $\langle \cdots\rangle_R$ is a steady-state  average over particle pairs conditional on their centre-of-mass distance $R$.  
 The order $p$ of these moments need not be an integer, and it can also
assume negative values. We consider small separations $R$ between the  particles, in the dissipation range of turbulence. 
 In this range  the second-order longitudinal velocity structure function $ \langle \delta u_L^2\rangle_R$ of the turbulent flow scales as  \cite{Sto93,0295-5075-34-6-411,Che05,Fri97,2010Blum,2015Ni}
 \begin{equation}
 \label{eq:uL}
  \langle \delta u_L^2\rangle_R/u_{\rm K}^2 = \tfrac{1}{15} (R/\eta_{\rm K})^2\,.
  \end{equation}
   Here $\delta u_L = [\ve u(\ve x+\ve R,t)-\ve u(\ve x,t)]\cdot \hat{\ve R}$ is the velocity increment of the turbulent velocity in the direction $\hat{\ve R}$
   of separation $\ve R$. This means that the turbulent {\em velocity} field is smooth in the dissipation range. It can be Taylor-expanded to give
   $\delta u_L \propto R$.
 If the {\em director} field $\ve n(\ve x,t)$ were smooth too, then $S_p(R) \propto R^p$ for small $R$, as in Eq.~(\ref{eq:uL}). But our results
show that this is usually not the case. In our DNS and statistical-model simulations we find instead \lq anomalous scaling\rq{}
of the director  field
\begin{equation}
\label{eq:power_law}
S_p(R) \sim (R/\eta_{\rm K})^{\xi_p} \,\,\,\mbox{for}\,\,\, R/\eta_{\rm K} \ll 1\,, \,\,\,\mbox{with}\,\,\, |\xi_p| \ll |p|\,.
\end{equation} 
Since $|\xi_p|\ll |p|$ this implies that angles between the symmetry axes of nearby particles are anomalously enhanced at small separations. This means that the spatial director field of non-spherical particles is not smooth, in general, so that angles between the symmetry axes of nearby non-spherical particles in turbulence are larger than expected. This is our main result. The remainder of this paper is organised as follows. In Section \ref{app:method} we  give details about our experiments, DNS, and the statistical model that we use to analyse the director patterns. Section \ref{section:results} summarises the results of
our experiments, DNS, and statistical-model analysis.  Section \ref{sec:conc} contains discussion, conclusions, and an outlook.

\begin{figure}[t]
\raisebox{1cm}{\begin{overpic}[width=7cm]{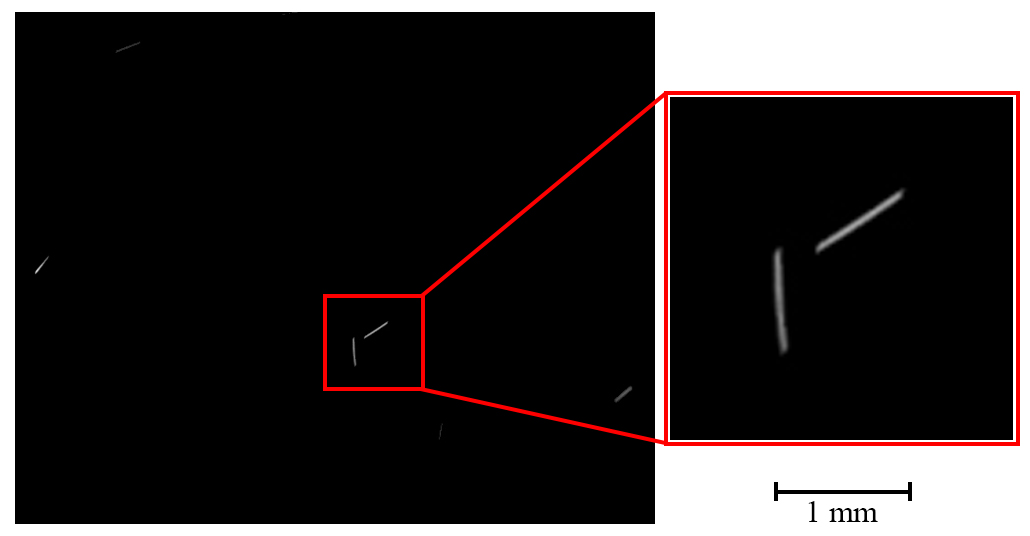}
\put(1.1,48){\colorbox{white}{\bf a}}
\end{overpic}}\hspace*{8mm}
\begin{overpic}[width=6.cm]{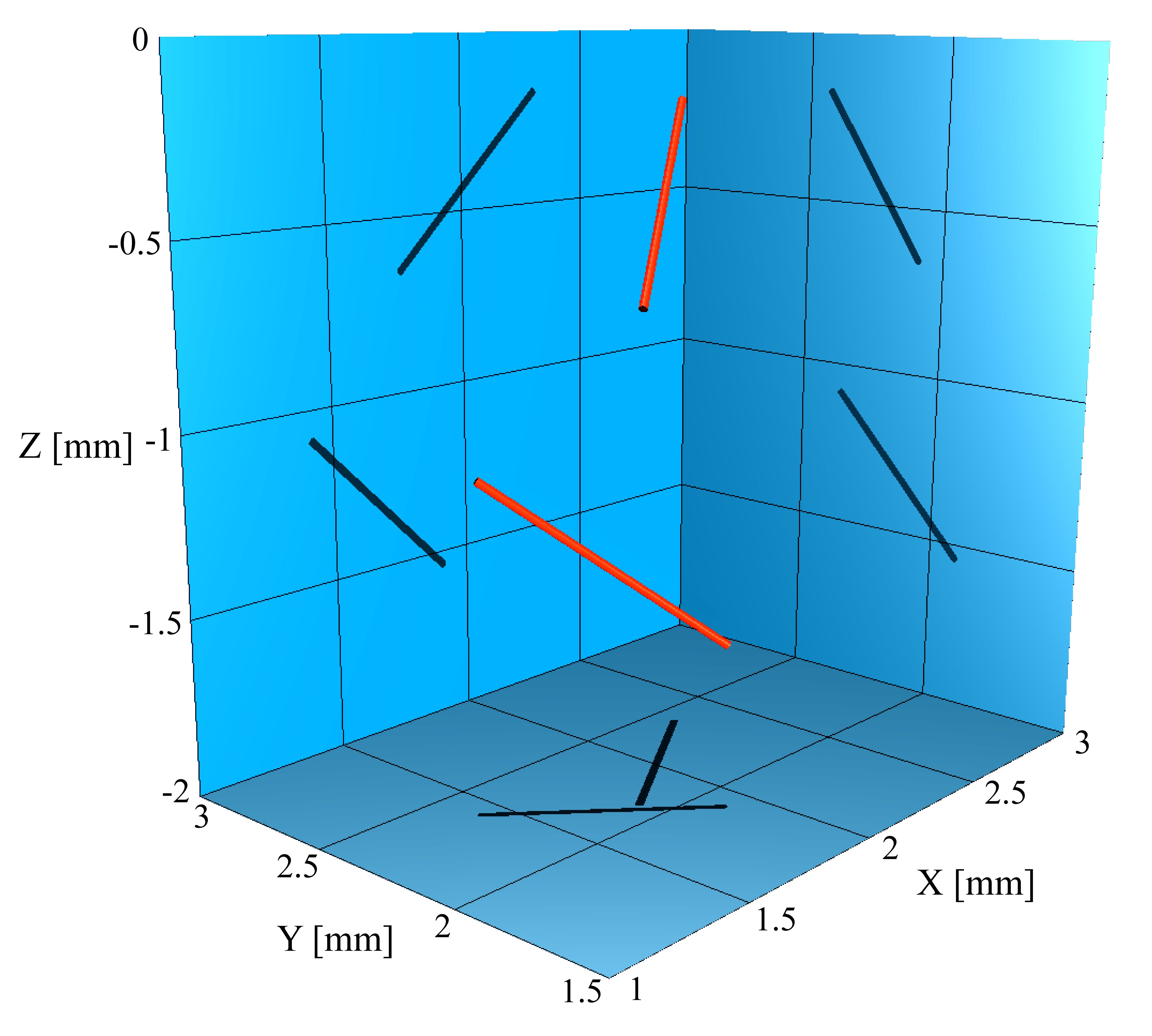}
\put(-6.,50){\colorbox{white}{\tiny $x_3$ [mm]}}
\put(19.,7.5){\colorbox{white}{\tiny $x_2$ [mm]}}
\put(78.,14){\colorbox{white}{\tiny $x_1$ [mm]}}
\put(90,82){\colorbox{white}{\bf b}}
\end{overpic}
\caption{\label{fig:raw-image}
Analysis of experimental raw data.
({\bf a}) Image from one camera showing two nearby rods.  The distance
between the rods (center-to-center) is $1.27$ mm, about two
rod lengths.  Acquired at volume illumination with ${\rm Re}_\lambda=277$. 
({\bf b})  Reconstruction of the directors of two nearby rods
from the experimental data. Same pair as shown in panel ({\bf a})}.
\end{figure}
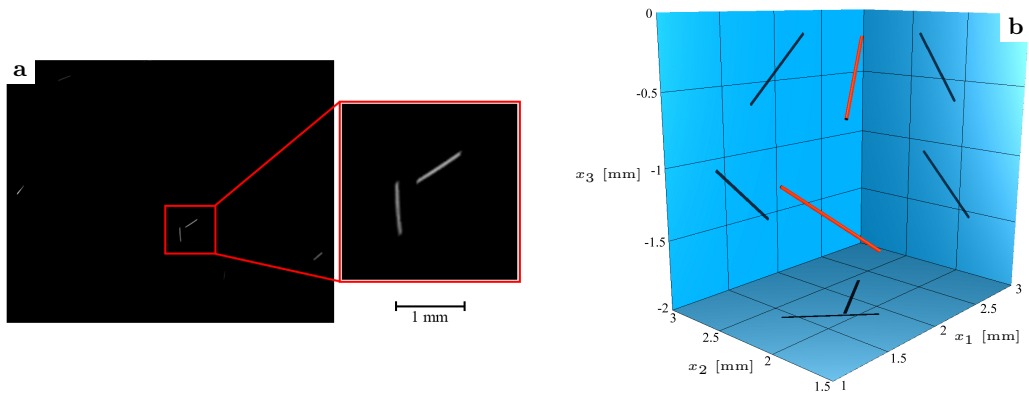

\section{Methods}
\label{app:method}
\subsection{Experiment}
\label{app:exp}
The experiments measure angle differences between rods near the centre of a
$1\times 1\times 1.5$\,m$^3$
octagonal tank between two oscillating grids \cite{2010Blum}.
Fluorescent dyed rods (length $a=700\,\mu$m, diameter $b=30\,\mu$m, aspect ratio $\kappa=23.3$)
are suspended in the flow. We estimate the particle-shape parameter $\Lambda$
by using Eq.~(\ref{eq:Lambda})
the formula for a spheroidal particle. This gives $\Lambda = 0.996$.
 Data is shown for two different grid frequencies, 1 and $3\,$Hz in water
with kinematic viscosity $\nu=0.96\times10^{-6}$ m$^2$ s$^{-1}$.
The resulting Taylor-scale Reynolds numbers are $\textrm{Re}_{\lambda}=140$ and $277$.
The energy dissipation rates of $\epsilon=9\times10^{-5}$ and $2.5\times10^{-3}$ m$^2$ s$^{-3}$ were calculated from the mean value of the compensated third-order structure function in the inertial range \cite{2010Blum,2015Ni}. The Kolmogorov lengths are $\eta_{\textrm{K}}\equiv(\nu^3/\epsilon)^{1/4}=310$ and $135$ $\mu$m,
and the Kolmogorov times are $\tau_{\textrm{K}}\equiv(\nu/\epsilon)^{1/2} =93$ and $19$ ms.

In the $\textrm{Re}_{\lambda}=140$-flow  the rod length is $2.3  \eta_{\textrm{K}}$ which is small enough that the particles are in the tracer limit. In the $\textrm{Re}_{\lambda}=277$-flow, the rod length is $5.2\eta_{\textrm{K}}$.
For particles of this size, tumbling rates are still roughly 
in the tracer limit \cite{Voth14},
but finite-size effects start to become important.  The Stokes numbers of the rods, defined as the ratio of the rod response time to the Kolmogorov time, are $0.002$ and $0.01$, so the particles behave like neutrally buoyant tracers even though the fluid density, $\rho_{\rm f} = 1.00$ g cm$^{-3}$, is slightly lower than the particle density, $\rho_{\textrm{p}}=1.12$ g cm$^{-3}$.

The two experiments used different imaging setups.  The lower Reynolds-number data was taken with a laser scanning system and three  cameras
recording 5000 frames per second (fps) \cite{2015Ni}. This dataset has 3.8$\times 10^4$ frames.
Each frame imaged one of eight slabs that were scanned sequentially.  A typical image contains $60$ rods within a slab with dimensions 1.5 cm $\times$ 1.5 cm $\times$ 0.3 cm.   The higher Reynolds-number data was taken with the imaging system using volume illumination and four cameras
at 450 fps \cite{Voth14}. This dataset has 1.5$\times10^6$ frames each with typically 8  rods in view in an imaging volume with dimensions 2 cm $\times$ 2 cm $\times$ 3 cm.  The seeding densities chosen are a compromise between obtaining sufficient numbers of rod-angle differences  at small distances and minimising the overlap of rods in the 2D images.  When rods overlap, it is difficult to separate them and measure their 3D  positions. We discard such samples. Because rods with large angle differences are more likely to overlap in the images, this introduces a sampling bias at small $R$.   
We therefore only report data for $R$ greater than the rod length, where the bias is not large.

The experimental raw data are analysed as follows. The camera images [Fig.~\ref{fig:raw-image}({\bf a})] are first segmented, identifying clusters of bright pixels.    The two-dimensional centres-of-mass of clusters are then stereo-matched and tracked over time using the camera calibration data and a predictive tracking algorithm \cite{2006Ouellette}.  Rod angles are  extracted from multiple images using methods described previously \cite{Par12,2016Cole}.
Fig.~\ref{fig:raw-image}({\bf b}) shows the three-dimensional reconstruction of the locations and angles between a pair of nearby rods
[same as the pair shown in Fig.~\ref{fig:raw-image}({\bf a})].
The rods used in these experiments have an aspect ratio of more than four times that of previous experiments on tracer rods~\cite{Par12} resulting in a smaller  uncertainty of the relative angle, compared to previous measurements.

\subsection{Direct numerical simulations}
\label{app:dns}
We perform DNS of a turbulent channel flow using one-way coupling for
the particle dynamics: given the fluid-velocity field and its gradient, spheroids with shape parameter
$\Lambda$  move according to Eqs.~(\ref{eq:adv}) and (\ref{eq:jeffery}).
The turbulent channel flow is characterised by the
Reynolds number ${\rm Re}_* = u_* h /\nu$ based on the
wall-friction velocity $u_*$ and the half-channel height $h$.
The wall-friction velocity is determined by the wall stress and the fluid density.
Since the channel flow is inhomogeneous, ${\rm Re}_\lambda$ varies throughout the channel cross section.
We take our statistics near the channel centre, in a region of linear size $2\eta_{\rm K}$, where the turbulent
vorticity is approximately homogeneous and isotropic 
\cite{andersson_anisotropic_2015}. 
In this region we estimate ${\rm Re}_\lambda= u_{\rm rms}' \lambda / \nu$ using
the local rms turbulent velocity $u_{\rm rms}' = \langle |\ve u'|^2/3\rangle^{1/2}$ and
the Taylor scale $\lambda = u'_{\rm rms} \sqrt{15\nu/\epsilon}$. The prime denotes
the fluctuating part of the fluid velocity obtained by Reynolds decomposition.
The dissipation rate is calculated from the local turbulent velocity gradients,
$\epsilon =\nu\langle {\rm Tr} \ma {A'}^\T\ma A'\rangle$.er
We choose ${\rm Re}_\ast=180$. Near the channel centre this gives
$u^{'+}_{\rm rms} = 0.686$, $\epsilon^+ = 5.4\times 10^{-3}$,
$\lambda^+ = 36.2$, ${\rm Re}_\lambda = 24.8$, $\eta^+_{\rm K}= 3.68$, $\tau^+_{\rm K}=13.6$,
and $u^+_{\rm K}=0.27$. All non-dimensional quantities are quoted in wall units, expressed
in terms of $u_\ast$ and $\nu$.
The simulation domain is $12h \times 6h  \times 2h$
in the streamwise, spanwise, and wall-normal directions.
We apply periodic boundary conditions in the spanwise and
streamwise directions, and no-slip boundary conditions at the two walls.
We use a pseudo-spectral method in
the periodic directions, and a $2^{\rm nd}$-order central finite-difference scheme \cite{zhao_rotation_2015} in the wall-normal direction. For time integration we use an explicit $2^{\rm nd}$-order Adams-Bashforth scheme.

\subsection{Statistical model}
\label{app:statm}
In two spatial dimensions we use a stream function $\Psi(\ve x,t)$
to represent a smooth, incompressible, homogeneous, isotropic random Gaussian velocity field $\ve u$:
we take \cite{Gus16a}
\begin{align}
\ve u=\tfrac{1}{\sqrt{2}}
\big[{\partial_y\Psi}\,,-{\partial_x\Psi} \big]^\T\,.
\end{align}
The stream function is constructed as a superposition of Fourier modes with Gaussian random time-dependent coefficients.  
The coefficients are chosen such that $\Psi(\ve x,t)$  has zero mean and correlations
\begin{align}
\langle\Psi(\ve x,t)\Psi(\ve x',t')\rangle=\eta^2u_0^2\exp\Big(-\frac{|\ve x-\ve x'|{^2}}{2\eta^2}-\frac{t-t'}{\tau}\Big)\,.
\eqnlab{statistical_model_corr_fun}
\end{align}
This correlation function defines the Eulerian scales of the flow, namely
the correlation length $\eta$ and the correlation time $\tau$. The typical speed
is $u_0\equiv\sqrt{\langle\ve u^2\rangle}$. 

In three spatial dimensions, the velocity field $\ve u(\ve x,t)$ is constructed as the rotation of a vector field    
$\ve A(\ve x,t)$ \cite{Gus16a}:
\begin{equation}
\ve u =  \frac{1}{\sqrt{6}} \ve \nabla \wedge \ve A(\ve x,t)\,.
\end{equation}
The three components  of $\ve A(\ve x,t)$ are independent Gaussian random functions with the same statistics as $\Psi(\ve x,t)$.
Further details of the statistical model are described in Ref.~\cite{Gus16a}.

\begin{figure}[t]
\begin{overpic}[width=\columnwidth]{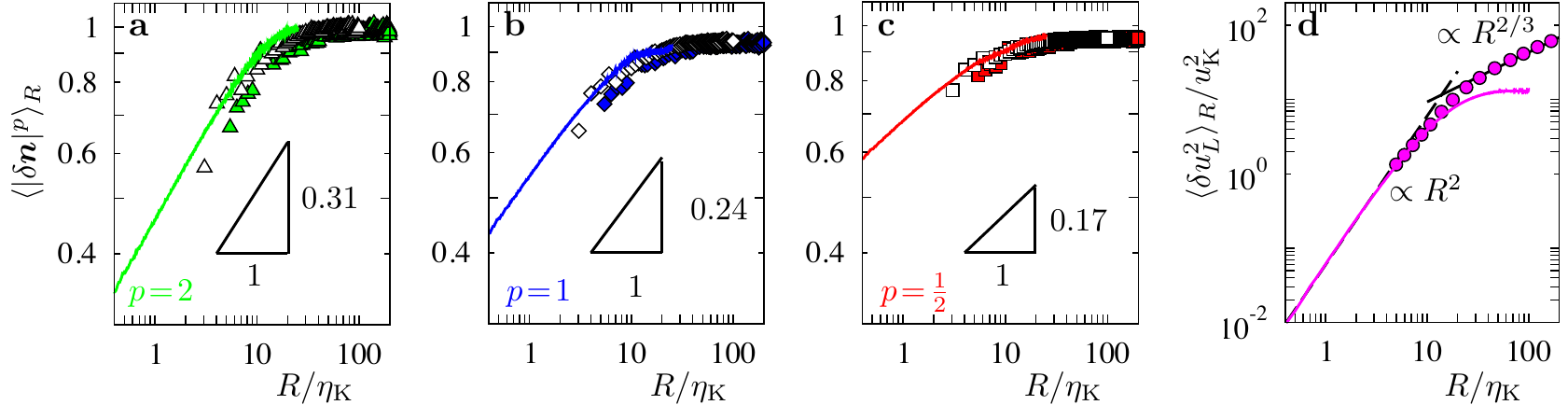}
\end{overpic}
\caption{\label{fig:1}  ({\bf a}) to ({\bf c}) Moments $\langle |\delta\ve n|^p\rangle_R$ of differences $\delta\ve n$ between directors  $\ve n$ of nearby non-spherical particles in turbulence, versus their centre-of-mass separations $R$.  Experimental results for rods (length $700\,\mu$m, diameter $30\,\mu$m) in a turbulent water tank  (Section \ref{app:method}), ${\rm Re}_\lambda = 277$, Kolmogorov length $\eta_{\rm K}=135\,\mu$m  (filled symbols), ${\rm Re}_\lambda = 140$, $\eta_{\rm K}=310\,\mu$m (open symbols).  
Direct-numerical simulation (DNS) results for spheroids with aspect ratio $\kappa\!=\!23.3$ in turbulent channel flow (Section \ref{app:method}), ${\rm Re}_\lambda=24.8$ (solid lines).  Small-$R$ fits  of  $(R/\eta_{\rm K})^{\xi_p}$ to the DNS data: $\xi_2\!=\!0.31$, $\xi_1\!=\!0.24$, $\xi_{0.5}\!=\!0.17$.  
 The data analysis is described in Section \ref{app:method}. 
 ({\bf d}) Longitudinal velocity structure function. The experimental data  is 
 taken from Fig.~4 in Ref.~\cite{2010Blum}, obtained for the same setup as this work (${\rm Re}_\lambda = 285$), symbols. DNS (this work), solid violet line. Also shown is
 the small-$R$ asymptote (\ref{eq:uL}) in the dissipation range (dashed black line), and in the inertial range where the scaling is approximately $R^{2/3}$ (solid black line).
}
\end{figure}

\section{Results}
\label{section:results}

This Section is organised in two parts: a summary of the experimental and DNS data (Section \ref{sec:as}), and
a summary of the statistical-model results (Section \ref{sec:sma}). 

In Section \ref{sec:as} we show our results for the angular structure functions $S_p(R)$  that illustrate our key conclusion, that the probability of observing large angles between nearby particles is anomalously large. Then we describe our analysis of the anomalous exponents $\xi_p$ in Eq.~(\ref{eq:power_law})  -- for DNS and statistical-model simulations in two and three spatial dimensions. 
The statistical-model  analysis (Section \ref{sec:sma}) yields three main results. First, the exponents $\xi_p$ saturate for large $p$
because the distribution of relative angles has power-law tails (Section  \ref{sec:xi_p}). Second, these tails result from a steady-state distribution of {\em steps} of different widths in the director  patterns that occur when turbulent strains act on particles aligned orthogonal to the main stretching direction (Section \ref{sec:scars}). We discuss how these steps are related to scar-line singularities observed in 2D random \cite{Wil09}, deterministic chaotic \cite{Voth17}, and quasi-periodic \cite{Wil11} maps. Third, we
show that  the attractor determining the steady-state director  patterns is fractal for small $\Lambda$, 
and thus not smooth (Section \ref{sec:fractal}). There is a phase transition: for  $\Lambda < \Lambda_{\rm c}$ the director  patterns are fractal, but for $\Lambda > \Lambda_{\rm c}$ they become locally smooth. In Section \ref{sec:3D} we demonstrate that  patterns
in  three spatial dimensions are similar to the patterns analysed for the two-dimensional statistical-model system.

\subsection{Anomalous scaling}
\label{sec:as}
\subsubsection{Angular structure functions}
Fig.~\ref{fig:1}({\bf a}) -- ({\bf c}) shows our experimental results for the angular structure functions $S_p(R)$ (symbols). In our experimental apparatus the dissipation range extends up to roughly  $10 \,\eta_{\rm K}$ where $\eta_{\rm K}$ is the Kolmogorov length. More specifically, Fig.~\ref{fig:1}({\bf d})
shows that  $\langle \delta u_L^2\rangle_R \propto R^2 $ in this range. The experimental data are taken from Fig.~4 in Ref.~\cite{2010Blum}, obtained
using the same measurement apparatus as in this paper, but at a slightly larger Reynolds number, ${\rm Re}_\lambda = 285$.  At larger separations
an inertial-range power law  \cite{Sto93,0295-5075-34-6-411,Che05,Fri97,2010Blum,2015Ni} emerges (the data are roughly consistent
with  $\langle \delta u_L^2\rangle_R \propto R^{2/3} $). The DNS data in Fig.~\ref{fig:1}({\bf d}) exhibit  dissipation-range scaling   $\langle \delta u_L^2\rangle_R \propto R^2$ up to approximately $10 \,\eta_{\rm K}$. There is no  inertial-range scaling for the DNS data because the Reynolds number is much lower, ${\rm Re}_\lambda = 24.8$.

 Panels ({\bf a}) to ({\bf c}) demonstrate
 that  the angular structure functions decay much more slowly than $R^p$ in the dissipation range. Also shown are results of our DNS (solid lines). 
 We see that experimental and DNS results agree in the range where we have both experimental and DNS data. In the experiment the range of spatial scales in the dissipative range is too small to extract reliable values for the scaling exponents, but the DNS results exhibit clear power-law scaling
 with anomalous exponents $\xi_p \ll p$ for $p= \tfrac{1}{2}, 1$, and $2$.

\subsubsection{Anomalous scaling exponents} 
\label{sec:anomalous}
\begin{figure}
\begin{overpic}[width=\columnwidth]{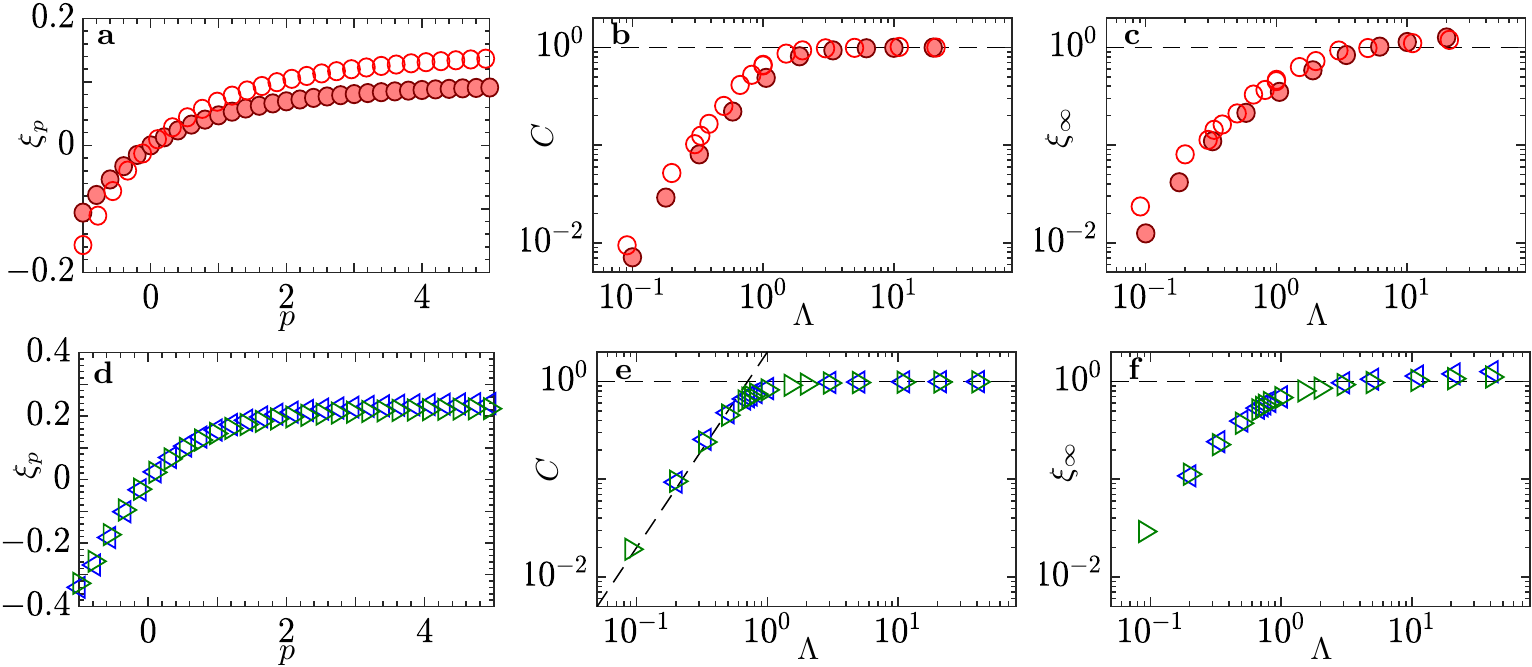}
\put(28,5.){\colorbox{white}{\small 2D}}
\put(62,5.){\colorbox{white}{\small 2D}}
\put(95.5,5.){\colorbox{white}{\small 2D}}
\put(28,27.){\colorbox{white}{\small 3D}}
\put(62,27.){\colorbox{white}{\small 3D}}
\put(95.5,27){\colorbox{white}{\small 3D}}
\put(40,7.){\colorbox{white}{\small $C(\Lambda)=2\Lambda^2$}}
\put(56,38){\small $C(\Lambda)=1$}
\put(56,15.5){\small $C(\Lambda)=1$}
\put(90,38){\small $\xi_\infty(\Lambda)=1$}
\put(90,15.5){\small $\xi_\infty(\Lambda)=1$}
\put(49,12){\vector(0,1){4}}
\put(48.3,10){\small $\Lambda_{\rm c}$}

\end{overpic}
\caption{\label{fig:dns}
Anomalous scaling exponents $\xi_p$ in Eq.~\eqnref{eq:power_law}.
({\bf a})  DNS results for $\xi_p$ versus $p$ for  $\Lambda=0.3$ ($\bullet$, red).
Also shown are three-dimensional (3D) statistical-model results for $\Lambda=0.3$ and $\ku=10$ ($\circ$).
({\bf b}) Results of fits to Eq.~\eqnref{eq:smallp}, $\xi_p = C(\Lambda)\,p$ for small $|p|$.  DNS ($\bullet$, red), 3D statistical model ($\circ$).
({\bf c}) Plateau value $\xi_\infty(\Lambda)$, Eq.~\eqnref{eq:largep}, versus $\Lambda$.
DNS ($\bullet$, red), 3D statistical model ($\circ$, red).
({\bf d})-({\bf f})  Same as panels {\bf a}--{\bf c} but for two-dimensional (2D) statistical-model simulations with $\Lambda=0.33$, $\ku=0.1$ ($\vartriangleright$, green), and $\ku=1$  ($\vartriangleleft$, blue).
The parameter $\Lambda_{\rm c}$ is defined in the text below Eq.~\eqnref{eq:result_DL}.}
\end{figure}

Fig.~\ref{fig:dns}({\bf a}) to ({\bf c}) shows DNS results for the  exponents $\xi_p$, as well as results of three-dimensional (3D) 
statistical-model simulations.  The DNS results shown in Fig.~\ref{fig:dns} ({\bf a}) are obtained by fitting the DNS data for $\log S_p(R)$ to  $A+\xi_p \log R$. We consider two fitting ranges, $0.05 \leq R/\eta_{\rm K} \leq 0.5$ and $0.02 \leq R/\eta_{\rm K} \leq 0.2$. The resulting estimates for $\xi_p$ are almost the same, the largest discrepancy is $5\%$. The data displayed in Fig.~\ref{fig:dns}({\bf a}) corresponds to $0.05 \leq R/\eta_{\rm K} \leq 0.5$.
The statistical-model results are obtained in a similar way. The values of $\xi_p$  are obtained by a linear least-squares fit  to $\log S_p(R)=A+\xi_p\log R$.  The data is fitted in two ranges: $0.005\eta<R<0.01\eta$ and $0.01\eta<R<0.05\eta$.  In both ranges the values of $C(\Lambda)$ and $\xi_\infty$ converge to the same values, except for $\Lambda\sim 1$ where differences of order $0.1$ are observed in $\xi_p$.  The data displayed is for the range $0.01\eta<R<0.05\eta$.

Fig.~\ref{fig:dns}({\bf a}) exhibits good qualitative agreement between 
these simulations.
For small values of $|p|$, the exponent $\xi_p$ is proportional to $p$ 
\begin{equation}
\label{eq:smallp}
\xi_p = C(\Lambda)\,p\quad\mbox{as}\quad p\to 0\,.
\end{equation}
To estimate the constant of proportionality, we fit $\xi_p=C(\Lambda) p$ to the DNS data in the range $-0.1 \leq p \leq  0.1$.
The corresponding statistical-model results are obtained from the average value of $\xi_p/p$ in the same range, $-0.1<p<0.1$.
Panel ({\bf b})  shows the results, how $C(\Lambda)$ depends on $\Lambda$.  
The values observed in the DNS and 
the statistical-model simulations are slightly different, but in both cases 
we find $C(\Lambda)\ll 1$  for small values of $\Lambda$.
 This observation is explained by the fact that the director  patterns are fractal, as we show below, and this is one source of the large differences between angles of nearby rods.  
 As $\Lambda$ increases, $C(\Lambda)$ grows. At large values of $\Lambda$ we see that  $C(\Lambda)$ approaches unity. 
 But as long as $C(\Lambda) < 1$ the  director field is fractal.
 
For large values of $p$, by contrast, $\xi_p$ tends to a constant. Both the DNS and the statistical-model
simulation show this behavior,  characteristic of highly intermittent fields~\cite{Fri97}
\begin{equation}
\label{eq:largep}
\xi_p \to \xi_\infty(\Lambda) \quad \mbox{as}\quad p \to \infty\,.
\end{equation}
We demonstrate below that this saturation is caused by narrow steps in the director   field across which  rods rotate by $\pi$. This is a second source of large         
   angle  differences at small $R$.  
Panel  ({\bf c}) shows how $\xi_\infty(\Lambda)$ depends on $\Lambda$.
The plateau values $\xi_\infty(\Lambda)$  are obtained as averages of $\xi_p$ in the interval $4<p<5$.

For large $\Lambda$, $\xi_\infty(\Lambda)$ is very close to unity. In summary, the scaling exponents are well approximated by $\xi_p \approx \min\{p,1\}$ in the limit of large $\Lambda$.

Note that the range of $\Lambda$ in panels  ({\bf b}) and  ({\bf c}) exceeds the physical limit  for slender rods, $\Lambda=1$. Equation (\ref{eq:Lambda}) shows that values of $\Lambda >1$ correspond to spheroids with  imaginary aspect ratios $\kappa$ (however one can construct particles that have $\Lambda >1$ \cite{Bretherton:1962}).
It is nevertheless instructive to consider the limit of large $\Lambda$:
we show in Section \ref{sec:xi_p} that the problem admits an exact solution for $\xi_p$  in the limit $\Lambda \gg 1$ and $\Lambda \ku\ll 1$. We discuss which insights this solution gives, and how it fails at small values of $\Lambda$.

 \subsection{Statistical-model analysis}
 \label{sec:sma}
 \subsubsection{Large-$\Lambda$ limit} 
 \label{sec:xi_p}
Figs.~\ref{fig:dns}({\bf d}) to ({\bf f}) show results of 2D statistical-model simulations. The results are very similar to the 3D case. We therefore begin by analysing the 2D model. The problem is simplest to analyse in the white-noise limit $\ku\to 0$, although turbulence corresponds to large Kubo numbers \cite{Gus16a}. We shall argue in this Section that this limit nevertheless yields important insights, just like  white-noise approximations for heavy-particle dynamics in turbulence \cite{Gus16a}, or Kraichnan's diffusion model for passive-scalar advection \cite{Fal01}.

\begin{figure}[t]
\begin{overpic}[width=12cm]{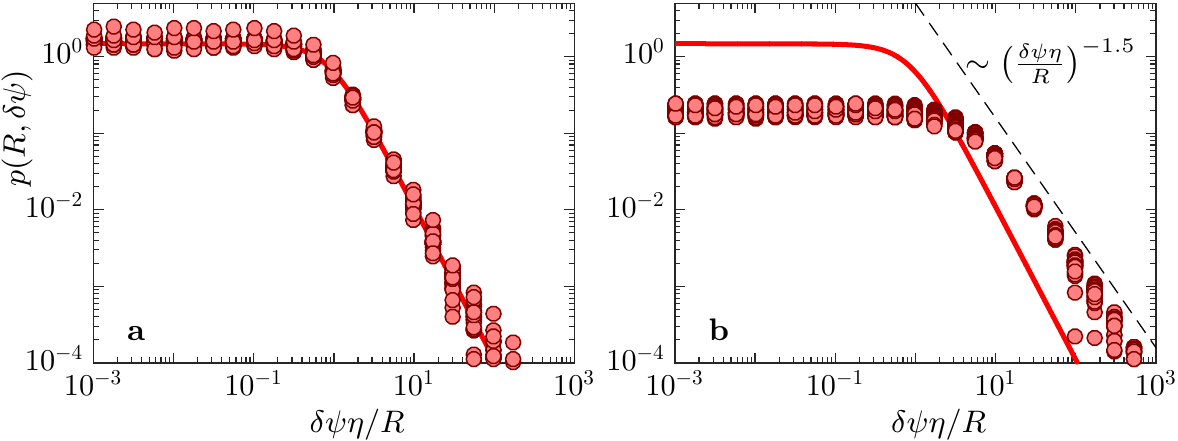}
\put(-1,21){\rotatebox{90}{\colorbox{white}{\small $P(R,\delta\psi)$}}}
\end{overpic}
\caption{\label{fig:distribution_sm}
Joint distribution  $P(R,\delta\psi)$
of centre-of-mass distance $R$ and relative
angle $\delta\psi$ in two-dimensional statistical model
for $\ku=0.01$ and $0.002\leq R/\eta\leq 0.08$.
({\bf a}) Simulation results ($\circ$, red)
for $\Lambda=21$.
Also shown is the large-$\Lambda$ theory, Eq.~\eqnref{eq:pR}, red solid line.
({\bf b}) Same, but for $\Lambda=2/3$. The slope of the power-law tail
in the numerical data is approximately $-1.5$,
different from the slope predicted by the
large-$\Lambda$ theory (red solid line), which is $-2$. 
For still smaller values of $\Lambda$ (not shown), numerical
results indicate that the distribution depends on $R$ and $\delta\psi$
separately, not only through $\delta\psi/R$.}
\end{figure}

The reason why the white-noise limit is simpler to analyse is that diffusion approximations can be applied. To make progress in our problem we must currently assume that  $\Lambda \gg 1$ and $\Lambda \ku\ll 1$. This means essentially that strain dominates over vorticity in aligning the fibres. In this limit we can compute the 
steady-state form of the 
distribution $P(R,\delta\psi)$ of centre-of-mass distances $R$ and angle differences $\delta\psi$ between particle pairs. Details
are given in Appendix \ref{app:PR}. The result is of power-law form:
\begin{equation}
\label{eq:pR}
P(R,\delta\psi) = \mathscr{N}/[1+\tfrac{4}{3} \delta\psi^2/R^2]\,.
\end{equation}
The factor  $\mathscr{N}$ is a normalisation constant.  
Fig.~\ref{fig:distribution_sm}({\bf a}) shows results of statistical-model simulations
at $\ku=0.01$ and $\Lambda=21$ for the joint distribution $P(R,\delta\psi)$ of angle differences
and separations. We see that Eq.~\eqnref{eq:pR} is an excellent
approximation at small Kubo numbers and large values of $\Lambda$.
Evaluating Eq.~\eqnref{eq:spr} with the distribution \eqnref{eq:pR} gives
\begin{equation}
\label{eq:spr_largeL}
S_p(R) \sim \langle |\delta\psi|^p\rangle_R
\sim
\left\{
\begin{array}{ll}
a_p R^p & \mbox{ for }p<1\,, \cr
b_p R & \mbox{ for }p>1\,,
\end{array}
\right.
\end{equation}
for small $R$ and $p\neq 1$, with coefficients $a_p = 2^{-p} 3^{p/2}\cos (p\pi/2)$ and $b_p =2^{-p+1} \sqrt{3} \pi^{p-2}/(p-1)$ (Appendix~\ref{app:PR}).  
Comparison with Eq.~(\ref{eq:power_law}) shows that
\begin{equation}
\label{eq:xipminp}
\xi_p = \min\{p,1\}
\end{equation}
for $\Lambda \gg 1$ and $\Lambda \ku\ll 1$. The saturation of $\xi_p$ for large $p$ is a consequence of the power-law tail of $P(R,\delta\psi)$.
Equation (\ref{eq:xipminp}) is consistent with the large-$\Lambda$ numerical results: panels {\bf b}, {\bf c}, {\bf e} and {\bf f} of Fig.~\ref{fig:dns} show that $C\approx 1$ and $\xi_\infty\approx 1$
down to $\Lambda =O(1)$. This indicates that the director  patterns are smooth for $\Lambda \gg 1$ and $\Lambda \ku\ll 1$. This is no longer true for small $\Lambda$.
In Section \ref{sec:fractal} we show that the director  patterns become fractal at small
values of $\Lambda$.

The large-$\Lambda$ approximation discussed above fails to account for the fractality observed at small $\Lambda$. However, numerical simulations demonstrate that the qualitative conclusions remain unchanged. Fig.~\ref{fig:distribution_sm}({\bf b})  shows that the distribution $P(R,\delta\psi)$ still has power-law tails,
albeit now with exponents different from $-2$ (the exponent is $\approx -1.5$ for $\lambda=\tfrac{2}{3}$). For small values of $R$, these power-law tails in $\delta\psi$ are cut off at $\delta\psi \sim \pi/2$, independent of $R$. As a consequence  the exponents $\xi_p$ saturate for  $p\gg 1$, but now at a constant smaller than unity.

\subsubsection{Scar lines} 
\label{sec:scars}
What causes the power-law tails in $P(R,\delta\psi)$?  
Consider a simple model in which strain is constant in space and time and vorticity is zero. We take
$\ma A = [[ -s, 0],[0,s]]$ and write $\ve n(\ve x,t) = [\cos\psi(\ve x,t),\sin\psi(\ve x,t)]$.  Integrating Eqs.~(\ref{eq:adv},\ref{eq:jeffery}) yields
\begin{equation}
\label{eq:tan}
\tan \psi(\ve x_t,t) = \exp(2\Lambda st) \tan \psi(\ve x_0,0)\,.
\end{equation}
Thus all initial angles converge toward the extensional strain direction, $\psi=\pm \pi/2$, except for $\psi(\ve x_0,0)=0$,
marking the location of steps of height $\pi$ in the director   pattern.
These steps are related to singularities observed in computer simulations of slender rods rotated by 2D random \cite{Wil09}, deterministic chaotic \cite{Voth17}, and quasi-periodic \cite{Wil11} maps.
These singularities occur where the extensional eigenvector of $\ma S$ is orthogonal to the initial  director pattern $\ve n(\ve x_0,0)$ \cite{Wil09,Fou12}, just as the steps in the example above.  In 2D, this constraint is satisfied on lines \cite{Wil09,Wil11,Voth17}, termed {\em scar lines} in Ref.~\cite{Wil09}.

In turbulence $\ma A(\ve x,t)$ changes as a function of space and time,
so that new steps are continuously created, old steps sharpen, and their height approaches $\pi$.  Since the problem is invariant under $\psi\to \psi+\pi$, the steps effectively disappear as they sharpen.
This is illustrated in Fig.~\ref{fig:sheets}({\bf a}) for $\Lambda=1$
where the attractor is smooth. Older steps leave thinner traces because they are less likely to be sampled by particles.  We conclude that a steady-state distribution of steps of different widths develops, independent of the initial condition.

How are these steps related to the power-law tails in $P(R,\delta\psi)$ and the saturation
of the exponents $\xi_p$ at large $p$? For large $\Lambda$, where the director   patterns are smooth, we can estimate the width $w_{\rm s}$ of a step in the $x_1$-direction as $w_{\rm s} \sim \pi/|\partial_{1}\psi|$, where $\partial_1\psi$ is the derivative of $\psi$ with respect to $x_1$.
In the limit of large $\Lambda$ and small $\Lambda \ku$ we find using diffusion approximations (Appendix \ref{app:Y1})  that $P(\partial_{1}\psi) \sim |\partial_{1}\psi|^{-2}$ for large values of $|\partial_{1}\psi|$.
To obtain the step-contributions to the angular structure function for large $p$, we note that a step of width $w_{\rm s}$ contributes with weight $w_{\rm s}/R$ to $\langle |\delta\psi|^p\rangle_R$, for $w_{\rm s}< 2R$. Wide steps with $w_{\rm s}> 2R$ give a smooth contribution of order $(R/w_{\rm s})^{p-1}$. Upon integrating over the distribution of $w_{\rm s}$ up to $w_{\rm s}\leq 2R$ we find that $\langle |\delta\psi|^p\rangle_R\sim R$ for large $p$, establishing the connection to Eq.~\eqnref{eq:spr_largeL}.  

\subsubsection{Fractal director patterns} 
\label{sec:fractal}
\begin{figure}[t]
 \begin{overpic}[width=6cm]{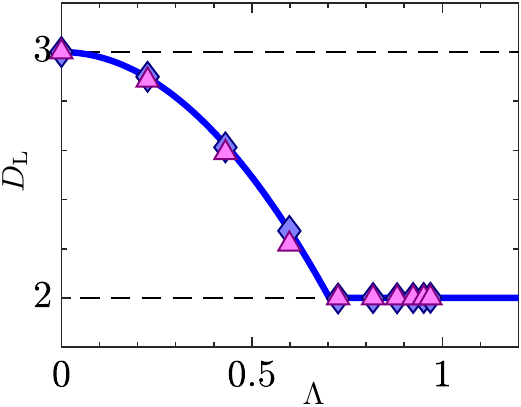}
 \end{overpic}
\caption{\label{fig:fractal_sm}
Lyapunov dimension $D_{\rm L}$ from two-dimensional statistical model simulations with $\ku=0.1$ and $\ku=1$, as a function of the shape parameter $\Lambda$.
$D_{\rm L}$ is shown as (\marker{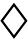},blue) for $\ku=0.1$, and as (\marker{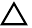},magenta) for $\ku=1$.
The theory \eqnref{eq:result_DL} -- evaluated with $\ku=0.1$ --  is shown as a solid blue line.
}
\end{figure}

The discussion in Sections \ref{sec:xi_p} and \ref{sec:scars} applies only 
in the limit of large $\Lambda$ when the    director patterns are smooth. 
Now we show that the director   patterns are fractal at smaller $\Lambda$. Consider first 2D. In this case the phase space of Eqs.~(\ref{eq:adv},\ref{eq:jeffery}) is three dimensional, spanned by the two components $x_1$ and $x_2$ of the centre-of-mass position $\ve x$, and by the 
angle $\psi$ of $\ve n$ with the $x_1$-axis, say.
We find that the scatter of points in this phase space is not smooth, but fractal. To characterise the fractal we
compute the Lyapunov dimension $D_{\rm L}$ \cite{Som93,Gus16a} 
to order $\ku^2$   (Appendix \ref{app:DL}). The result is:
\begin{equation}
D_{\rm L} =
\left\{
\begin{array}{ll}
3- 2\Lambda^2 + 4\ku^2\Lambda^2(\Lambda^2-1)\hspace*{-2mm} & \mbox{ for } \Lambda<\Lambda_{\rm c}\,,\cr
2 & \mbox{ for }\Lambda\ge \Lambda_{\rm c}\,,\cr
\end{array}
\right.
\label{eq:result_DL}
\end{equation}
with $\Lambda_{\rm c}\!=\!\tfrac{1}{\sqrt{2}}(1 -\tfrac{1}{2}\ku^2)$.
Eq.~\eqnref{eq:result_DL}  shows that the phase-space attractor is fractal
for $0 < \Lambda< \Lambda_{\rm c}$ because $D_{\rm L}$ is not an integer
in this range.  There is a {\em phase transition} at $\Lambda_{\rm c}$,
 the fractal dimension $D_{\rm L}$ equals two for $\Lambda > \Lambda_{\rm c}$, indicating that the attractor is smooth in this range.
 Fig.~\ref{fig:fractal_sm} shows numerical results from statistical-model simulations for $D_{\rm L}$, in two spatial dimensions.
We observe good agreement with Eq.~\eqnref{eq:result_DL} for small $\ku$.
We also see that $D_{\rm L}$ depends only very weakly on $\ku$ (note that Eq.~\eqnref{eq:result_DL}  does not
apply for $\ku=1$).
This indicates that preferential-sampling effects~\cite{Gus16a}  are weak.

For a Gaussian random function $f(x)$ with power-law spatial correlations, Orey \cite{Orey} derived a relation between the increments $\delta f\equiv f(\ve x+\ve R)-f(\ve x)$ and the fractal Hausdorff dimension $D$ of the set of points embedded in the $d+1$-dimensional space with coordinates $\ve x$  and $f$:
 \begin{equation}
 \label{eq:hurst}
 \langle |\delta f|^p\rangle_R \sim  R^{p(d+1-D)}\,.
 \end{equation}
To lowest order in $\ku$, Eq.~\eqnref{eq:result_DL} gives $D_{\rm L} \sim 3-2\Lambda^2$.
Setting $D=D_{\rm L}$ in Eq.~\eqnref{eq:hurst} yields $C(\Lambda) = 2\Lambda^2$, roughly consistent with the numerical results in Fig.~\ref{fig:dns}({\bf e}).  
Fig.~\ref{fig:sheets}({\bf b}) illustrates steps in the director   patterns for $\Lambda = 2/3$ where the attractor is fractal. Step-like structures are still present, but less distinct because
the attractor is fractal. 

In inviscid one-dimensional Burgers turbulence \cite{Bec:2007,Cha10} the turbulent velocity structure functions $\langle |\delta u|^p\rangle_R$  obey $\langle |\delta u|^p\rangle_R\sim R^{{{\zeta}}_p}$ at small $R$, and $\zeta_p\to 1$ at large $p$. Bifractal theory \cite{Fri97} relates the saturation of the exponent $\zeta_p$ at large $p$   to steps ({\em shocks}) in the velocity field.
This provides an insightful analogy: the saturation of the scaling exponents is caused by steps in the spatial field, as in our problem.
Yet there are fundamental differences: the velocity field in inviscid Burgers turbulence exhibits sharp jumps on a fractal set of positions, but the director field is itself fractal, in addition to exhibiting jumps.  How to calculate the exponents in our system for  $0 < \Lambda < 1$ is an open problem, even in the diffusive  limit $\ku\to 0$.  

\subsubsection{Three spatial dimensions} 
\label{sec:3D}
\begin{figure}
\begin{overpic}[width=0.6\columnwidth]{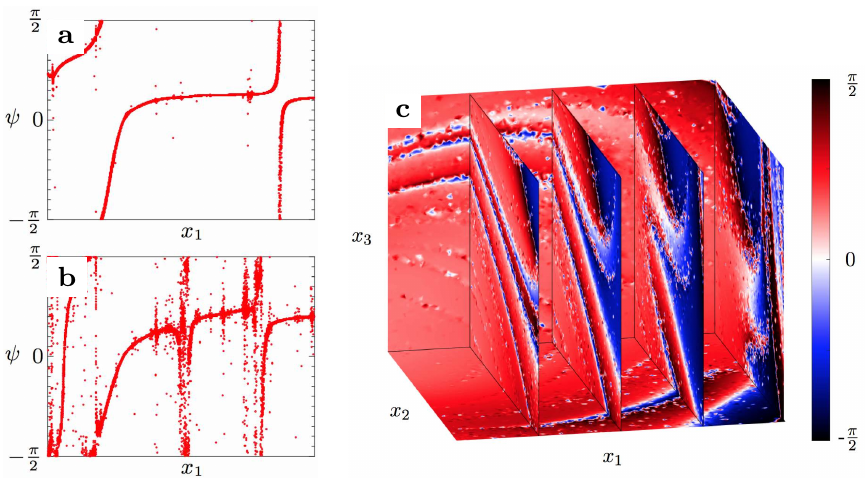}
\end{overpic}
\caption{\label{fig:sheets}
Steps in the director   field. ({\bf a}) Results of 2D statistical-model simulations. Angle $\psi$ of the director with the $x_1$-axis mod $\pi$, as a function of $x_1$ for a narrow range of $x_2$; $\Lambda=1$. ({\bf b})  Same, but for $\Lambda=2/3$.  In both panels $\ku=1$, the range of the $x_1$-axes is $0.7\eta$. Initial condition: random directors. Simulation time $20\, \tau$. Identical
fluid-velocity realisations for both panels.   ({\bf c})
Sheet-like steps in steady-state pattern of the azimuthal angle $\varphi$ from
3D statistical-model simulations.
Since $\varphi$ is defined mod $\pi$, the colour scheme
is wrapped in the same way.  Initially random directors. Size of the region shown:  $0.1\,\eta\times 0.1\,\eta\times 0.1\,\eta$. $\ku=1$ and $\Lambda = 1$. }
\end{figure}

The argument leading to Eq.~\eqnref{eq:tan} generalises to 3D.
We can therefore conclude that steps form also in
3D. In this case phase space is five-dimensional: three centre-of-mass dimensions
plus two Euler angles for the azimuthal ($\varphi$) and polar ($\theta$) degrees of freedom.  Fig.~\ref{fig:sheets}({\bf c}) shows a director   pattern from 3D statistical-model simulations. The value of $\varphi$ is colour-coded and plotted as a function of $x_1$, $x_2$, and $x_3$. The spatial pattern is consistent with steps:
in a given $x_2$-$x_3$-plane we see sharp transition lines where $\varphi$ jumps by $\pi$, and these lines appear also in neighbouring $x_2$-$x_3$-planes, at different values of $x_1$.

\begin{figure}[t]
 \begin{overpic}[width=6cm]{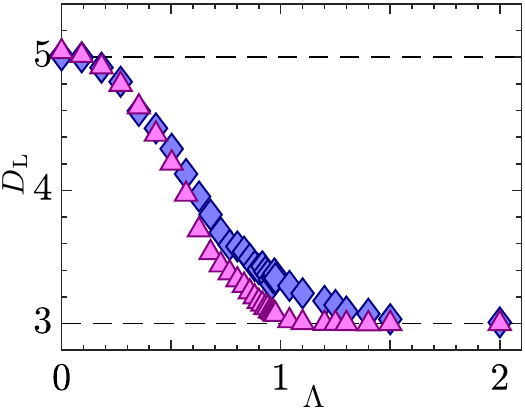}
 \put(92,72){\large \bf a}
 \end{overpic}
 \hspace*{8mm}
\raisebox{-1.6cm}{  \begin{overpic}[width=5.7cm]{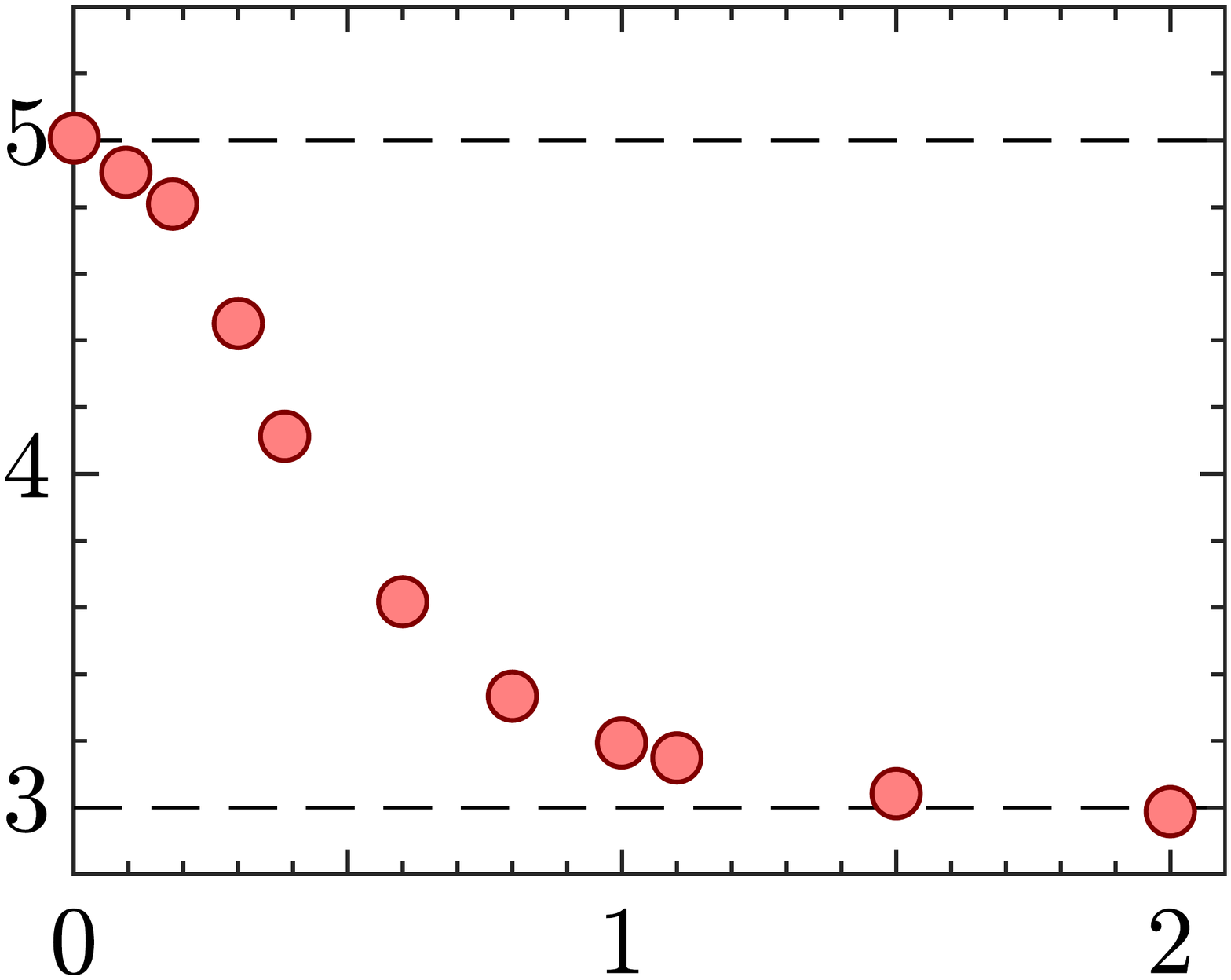}
\put(64,72){\large \bf b}
\put(-4.5,51){\rotatebox{90}{{\large $D_2$}}}
\put(37,21){ \large $\Lambda$}
\end{overpic}}
\caption{\label{fig:fractal_3D}
({\bf a}) Lyapunov dimension $D_{\rm L}$ from three-dimensional statistical model simulations with $\ku=0.1$ (\marker{3},blue) and $\ku=1$ (\marker{4},magenta), as a function of the shape parameter $\Lambda$.
({\bf b}) Correlation dimension $D_2$ ($\circ$, red) from DNS of a turbulent channel flow, as a function of the shape parameter.  
Values of $\Lambda$ larger than unity do not correspond  to physical rods (Section \ref{sec:anomalous}).
}
\end{figure}

For smaller  $\Lambda$ we expect that the director  patterns are fractal, as in 2D. The numerical results from 3D statistical-model simulations shown in Fig.~\ref{fig:fractal_3D}({\bf a}) demonstrate that this is the case. The results indicate that there is a phase transition, as in 2D. Since phase space is five-dimensional,  $D_{\rm L}$ changes from five
at $\Lambda=0$ to three at large values of $\Lambda$. For $\ku=1$ the critical
shape parameter is approximately $\Lambda_{\rm c}\approx 1$. 
Also shown is data for $\ku=0.1$. The results for
$\ku=1$ and $0.1$ are slightly different, unlike in two spatial dimensions.
This could be due to numerical errors: the $\ku=0.1$
data for $D_{\rm L}$ in three spatial
dimensions are the most difficult to obtain amongst the displayed
statistical-model data. But we cannot exclude
that there is a $\ku$-dependence in three spatial dimensions.
This would indicate that preferential effects  \cite{Gus16a} or large
time correlations matter.

For the turbulent channel flow our conclusions are qualitatively similar, although we  could not determine the Lyapunov dimension reliably because the flow is inhomogeneous, and the long trajectories needed to estimate $D_{\rm L}$ do not remain in the centre of the channel where we must take the statistics.  Therefore we have numerically computed the correlation dimension $D_2$. It is defined as \cite{Gra83}:
\begin{equation}
\label{eq:D2_def}
P(|\delta\ve w|) \sim |\delta\ve w|^{D_2-1}\quad\mbox{as}\quad |\delta\ve w|\to 0\,,
\end{equation}
where $\delta\ve w = [\delta x_1,\delta x_2,\delta x_3,\delta\theta, \delta \varphi]^\T$, and $\delta\theta$ and $\delta \varphi$ are differences of particle
azimuthal angles $\theta$ and polar angles $\varphi$.
The value of the power-law exponent in \eqnref{eq:D2_def} is obtained by fitting Eq.~(\ref{eq:D2_def}) to the DNS data, in the range
 $0.02< |\delta\ve w| <0.2$.  We also tested a slightly lower
range, from $0.01$ to $0.1$. This makes
only a small difference to the results, but to quantify the error it would be necessary to
measure at substantially smaller values of $|\delta\ve w|$. Our present DNS data does not permit this.

Fig.~\ref{fig:fractal_3D}({\bf b}) shows that the fractal correlation dimension       $D_2$   ranges from $5$ at $\Lambda =0$
to approximately $3$ at large values of $\Lambda$, like the Lyapunov dimension for the 3D statistical model.
However, for the DNS it is difficult to estimate $\Lambda_{\rm c}$ precisely.
Fig.~\ref{fig:fractal_3D}({\bf b})  indicates that the critical shape parameter is
 of order unity, so that the steady-state attractor is fractal for generic axisymmetric particles with fore-aft symmetry, except possibly for very slender rods.

\section{Discussion and conclusions.}
\label{sec:conc}
Figs.~\ref{fig:1}  shows experimental and DNS data that illustrate our key result: angles between the symmetry axes of nearby particles are anomalously large. 
We quantified this phenomenon by analysing angular structure functions $S_p(R)$ at small spatial scales $R$. These functions measure the moments of relative angles between nearby particles. DNS and statistical-model simulation data
demonstrate that the angular structure functions $S_p(R)$ exhibit power laws at small $R$, with anomalous scaling exponents $\xi_p$.
Using diffusion approximations in the limit $\Lambda\gg 1$ and $\Lambda\ku\ll 1$ we derived a theory that qualitatively explains the dependence of $\xi_p$ upon $p$. The exponents saturate at large $p$. This is  caused by turbulent strains that form a steady-state distribution of steps in the director  patterns. In the limit of $\Lambda\gg 1$ the director  patterns are spatially smooth. 
But for small $\Lambda$, the director  patterns are fractal,
and this explains the small-$p$ limit of the scaling exponents $\xi_p$. At large $p$ the exponents saturate, but
no longer to unity. The saturation is still a consequence of steps in the director    patterns, but now
 the steps inherit their properties from the fractal nature of the attractor.

The experimental angular structure functions agree very well with the DNS results (Fig.~\ref{fig:1}), down to separations of the order of the rod length, although the DNS is performed at much smaller Reynolds number. There is excellent agreement between the DNS results and our statistical-model predictions. These facts indicate that the mechanisms causing anomalously large angles are robust, so that our theory  provides a foundational framework for understanding the relative alignment of particles in turbulence.

Our problem is related to, yet fundamentally different from, passive-scalar and vector problems \cite{Pum94,Fri97,Shraiman,Fal01,Ott89,Che99,Fal01,Vin02}. 
Instead of the magnitude of a scalar or a vector, we analyse the spatial  director  field $\ve n$, normalised to unity and invariant under $\ve n \to -\ve n$.  
Moreover, we consider {\em finite} particle distances in the dissipative range of turbulence. This is a {\em two-particle} problem, more general than the question of how spatial gradients of angles evolve, advected by turbulence. The latter dynamics is a {\em single-particle} problem,  it refers to an initially smooth  manifold in phase space, as does the question of how the curvature of a material surface evolves in turbulence \cite{pope:1989}.

An open question is to find the form of the joint distribution $P(R,\delta\psi)$ of separations $R$ and angles $\delta\psi$ for finite  $\Lambda$. Our numerical results indicate that the distribution has algebraic tails too, and that the power-law exponents vary as a function of particle shape.
A related open problem is to derive the distribution in three spatial dimensions.  

In the experiment it is difficult to separate rods that overlap in the camera images.  
In Fig.~\ref{fig:1} we therefore show the experimental data only for distances larger than a rod length. It would be of great interest to obtain precise data at smaller distances, to systematically study the effect of hydrodynamic interactions  
\cite{shaqfeh,GuazzelliHinch} in the presence of non-trivial fluid flow.

A more far-reaching and more difficult problem is to understand how inertial effects change the patterns formed by larger particles. 
Particle inertia is relatively straightforward to take into account using the techniques reviewed in Ref.~\cite{Gus16a}. We anticipate that
caustics \cite{Fal02,Wil06,Bec10,Bewley} in the angular dynamics may  increase the probability of large angles between nearby particles.  
For larger particles it also matters how the particles accelerate the surrounding fluid. There are corrections due to turbulent shears \cite{Can16}, and convective fluid inertia must matter for rapidly settling particles.

The relative alignment of particles in turbulence is  a critical question in many scientific and engineering problems (including scattering of electromagnetic radiation from icy clouds  \cite{Pru78}, the dynamics of fibre suspensions \cite{Lun11}, and plankton ecology \cite{Gua12}). Specifically, the relative alignment of approaching particles
 affects the rate at which the they  collide, and possibly also collision outcomes. An ambitious long-term goal is to derive a theory for the collision rate between non-spherical particles in turbulence, with industrial fibre flows in mind \cite{Lun11}, but also to model encounter rates of motile microorganisms\cite{Gua12,Ber16} and of organic matter \cite{KH93}
 in the turbulent ocean. In this context we must also ask: how does breaking of fore-aft symmetry change the results summarised here?  A first step in this direction is to use the present theory to determine the statistics of differences between the angular velocities  $\ve \omega = \ve \Omega + \Lambda \ve n\wedge \ma S\ve n$ of nearby particles.  We expect that fractal steps in the director patterns are important, because the angular velocity depends explicitly on $\ve n$, and upon the shape parameter $\Lambda$.  But this is just a starting point. The general problem is a very difficult one, yet important because of its wide range of applications. Our results show that the analysis of statistical models is a promising way of approaching this impactful question.

\begin{acknowledgments} 
BM thanks S. \"Ostlund, J. Einarsson, J. Meibohm, and M. Cencini for discussions.  KG and BM were supported by the grant {\em Bottlenecks for particle growth in turbulent aerosols} from the Knut and Alice Wallenberg Foundation, Dnr. KAW 2014.0048, and in part by VR grants nos. 2013-3992 and 2017-3865. Computational resources were provided by C3SE and SNIC. LZ and HA benefited from a grant (project no. 250744/F20 'Turbulence-plankton interactions') and computational resources (grant no. NN2649K) from The Research Council of Norway. RN, SK and GV were supported by NSF grant DMR-1508575.
\end{acknowledgments}

\appendix

\section{Calculation of $P(R,\delta\psi)$ using diffusion approximations}
\label{app:PR}
Linearising the dynamics, Eqs.~(\ref{eq:adv},\ref{eq:jeffery}), for two nearby directors of separation $\ve R\equiv\ve x_2-\ve x_1$ and relative  angle $\delta\psi\equiv\psi_2-\psi_1$ we find in two spatial dimensions:
\begin{subequations}
\begin{align}
\ddt{\ve R}&={\ku\,}\ma A\ve R\\
\ddt\psi&
={\ku}\,\epsilon_{jk}\left[\tfrac{1}{2}O_{kj}-\Lambda n_kn_l S_{jl}\right]\,,\\
\ddt{\delta\psi}&
={\ku}\left[ n_j\epsilon_{jl}\partial_iB{_{lk}}R_in_k-2\Lambda \,\delta\psi\, n_jS_{jk}n_k\right]\,.
\end{align}
\eqnlab{sep_dynamics}
\end{subequations}
Here we have represented
$\ve n$ as $[\cos\psi,\sin\psi]^\T$, and we have expanded the
fluid velocity $\ve u$ and the angular velocity $\omega$ in terms of small separations. Eq.~(\ref{eq:sep_dynamics}) is expressed in dimensionless variables, $x'=x/\eta$, $t'=t/\tau$, $u' = u/u_0$, and we have
dropped the primes.
Further, $\ma B=\ma O+\Lambda\ma S$, repeated indices are summed
(Einstein summation convention), and $\epsilon_{jl}$ is the two-dimensional Levi-Civita symbol.  For small Kubo numbers, the fluid-velocity gradients and second derivatives in \Eqnref{eq:sep_dynamics} fluctuate rapidly and can be approximated by white noise. In this limit, we approximate the  dynamics \eqnref{eq:sep_dynamics} by a four-dimensional diffusion process. For the diffusion approximation to hold we must require not only that $\ku\to 0$ (white-noise limit), but also that the change in $\ve n$ during one correlation time $\tau$ is much smaller than the magnitude $|\ve n|=1$.
We find that this change is of order $\ku\Lambda\ve n\cdot \ma S\ve n$ for large $\Lambda$.
Since the magnitude of $\ve n\cdot \ma S\ve n$ is of order unity in the dimensionless variables adopted in \Eqnref{eq:sep_dynamics}, we must require that $|\Lambda|\ku\ll 1$ for the diffusion approximation to hold.

We calculate the drift and diffusion coefficients using an expansion
in the Kubo number \cite{Gus16a}.  We use the dimensionless variables in \Eqnref{eq:sep_dynamics}. We find that the drift coefficients vanish,
and that the diffusion coefficients are given by:
\begin{subequations}
\begin{align}
D_{R_iR_j}&=\frac{\ku^2}{2}(3\delta_{ij}R^2-2R_i R_j)\,,\\
D_{R_i\psi}&=-\frac{\ku^2}{2}\big[(\Lambda(\ve n\cdot\ve R)\epsilon_{ij}n_j + 2\epsilon_{ij}R_j+\Lambda R_j\epsilon_{jk}n_k n_i\big]\,,\\
D_{R_i\delta\psi}&=\ku^2\Lambda\delta\psi\big[R_i-2(\ve n\cdot\ve R)n_i\big]\,,\\
D_{\psi\psi}&=\frac{\ku^2}{2}(2+\Lambda^2)\,,\\
D_{\psi\delta\psi}&=0\,,\\
D_{\delta\psi\delta\psi}&=\frac{\ku^2}{2}\big[4\Lambda^2\delta\psi^2 + 12\Lambda(\ve n\cdot\ve R)^2 +(6-6\Lambda+3\Lambda^2)R^2\big]\,.
\end{align}
\end{subequations}
In the resulting diffusion equation, we
change to polar coordinates $\ve R\equiv R[\cos\beta,\sin\beta]^\T$.
Since the flow is isotropic, we can remove one angular degree of freedom, leaving a three-dimensional
diffusion equation in terms of $R$, $\delta\psi$, and $\delta\beta\equiv\beta-\psi$.
This three-dimensional equation is hard to solve in general, but in the limit of large $\Lambda$ we have found the solution.
Keeping only terms to highest order in $\Lambda$, the diffusion equation
simplifies considerably. In this limit, the steady-state diffusion equation reads:
\begin{align}
0&=8f
+ 16\delta\psi\partial_{\delta\psi}f
+(4\delta\psi^2 + 3R^2)\partial^2_{\delta\psi}f
+\partial^2_{\delta\beta}f\,.
\label{eq:diff_f}
\end{align}
Here $f$ is a function of $R$, $\delta\psi$ and $\delta\beta$. It is related to the probability distribution $P(R,\delta\psi,\delta\beta)$ by $f=P/R$.
To solve Eq.~\eqnref{eq:diff_f} we attempt a Fourier expansion
\begin{equation}
f(R,\delta\psi,\delta\beta)=\sum_m f_m(R,\delta\psi)e^{2\pi i m\,\delta\beta}\,.
\end{equation}
This ansatz results in:
\begin{align}
0&=(8-4 m^2 \pi^2)f_m
+ 16\delta\psi\partial_{\delta\psi}f_m+(4\delta\psi^2 + 3R^2)\partial^2_{\delta\psi}f_m\,.
\label{eq:diff_f_m}
\end{align}
The solution of this equation has power-law tails for large $\delta\psi/R$. The large-$\delta\psi/R$ asymptote of the solution can be written as a combination of two independent power laws in $\delta\psi/R$:
\begin{equation}
\label{eq:f_m}
f_m\sim a_m(R)\big(\tfrac{\delta\psi}{R}\big)^{-3/2 + 1/2 \sqrt{1 + 4 m^2 \pi^2}}+b_m(R)\big(\tfrac{\delta\psi}{R}\big)^{{\rm max}\{-3/2 - 1/2 \sqrt{1 + 4 m^2 \pi^2},-7/2 + 1/2 \sqrt{1 + 4 m^2 \pi^2}\}}\,.
\end{equation}
We must require that the tails are integrable to $|\delta\psi/R|=\infty$.
This implies that $a_m(R)=0$ for all values of $m$, and that $b_m(R)=0$ for $m\ne 0$.
Since the centre-of-mass of the particles is advected
in an incompressible flow, the marginal (spatial) distribution must be uniform.
Therefore we we must require that
\begin{align}
\int_0^{\pi/2}{\rm d}\delta\psi f_0(R,\delta\psi)=\frac{1}{R}\int_0^{\pi/2}{\rm d}\delta\psi P(R,\delta\psi)\sim {\rm const.}
\eqnlab{marginal_asymptote}
\end{align}
for small values of $R$.
We now match the general $m=0$-solution of \eqnref{eq:diff_f_m},
\begin{equation}
f_0(R,\delta\psi) = \frac{4}{3}\frac{a_0(R)\tfrac{\delta\psi}{R} +b_0(R)}{1+\tfrac{4}{3}\tfrac{\delta\psi^2}{R^2}}\,,
\end{equation}
to the asymptote \eqnref{eq:marginal_asymptote}.
As concluded above we have $a_0(R)=0$, so that we must match $b_0(R)=3\mathscr{N}/(4R)$ for small values of $R$, where $\mathscr{N}$ is a normalisation factor.
For the distribution $P=f_0R$ we thus obtain
\begin{equation}
\label{eq:supp_pR}
P(R,\delta\psi) = \mathscr{N}/[1+\tfrac{4}{3} \delta\psi^2/R^2]\,.
\end{equation}
This is Eq.~(\ref{eq:pR}).
Evaluating the moments $\langle |\delta\psi|^p\rangle_R$
with this distribution yields
\begin{equation}
\label{eq:supp_spr_largeL}
S_p(R) \sim \langle |\delta\psi|^p\rangle_R
\sim
\left\{
\begin{array}{ll}
a_p R^p & \mbox{ if }p<1 \cr
b_p R & \mbox{ if }p>1
\end{array}
\right.
\end{equation}
for small $R$ and $p\neq 1$.
This is  Eq.~(\ref{eq:spr_largeL}),
and the coefficients are given by
$a_p = 2^{-p} 3^{p/2}\cos (p\pi/2)$ and $b_p =2^{-p+1} \sqrt{3} \pi^{p-2}/(p-1)$ for $p\ne 1$.
For $p = 1$ we find logarithmic corrections to power-law scaling, $S_1(R)\sim R\log R$.

\section{Calculation of the distribution of $Y_1\equiv\partial_1\psi$}
\label{app:Y1}
Starting from Eqs.~(\ref{eq:adv},\ref{eq:jeffery}), the joint dynamics of angles and angle gradients $Y_i\equiv\partial_i\psi$ becomes in
two spatial dimensions:
\begin{subequations}
\begin{align}
\ddt\psi&=\ku\,\epsilon_{jk}\big(\tfrac{1}{2}O_{kj}-\Lambda n_kn_l S_{jl}\big)\,,\\
\ddt{Y_i}&=\ku\left[n_j\epsilon_{jl}B_{lk,i}n_k-2\Lambda Y_in_jS_{jk}n_k-Y_jA_{ji}\right]\,,
\end{align}
\eqnlab{gradient_dynamics}
\end{subequations}
using the same notation as in Section S-4.
In addition, $B_{lk,i}=\partial_i(O{_{lk}}+\Lambda S{_{lk}})$.
In the white-noise limit the dynamics \eqnref{eq:gradient_dynamics} describes a three-dimensional diffusion process with drift
\begin{subequations}
\begin{eqnarray}
D_\psi&=&0\,,\\
D_{Y_i}&=&2\ku^2\Lambda[2(\ve n\cdot\ve Y)n_i-Y_i]\,,
\end{eqnarray}
\end{subequations}
and with diffusion coefficients
\begin{subequations}
\begin{eqnarray}
D_{\psi\psi}&=&\ku^2\big(1+\tfrac{1}{2}\Lambda^2\big)\,,\\
D_{\psi Y_i}&=&\ku^2\big(-\epsilon_{ij}Y_j - \tfrac{1}{2}\Lambda n_in_j\epsilon_{jk}Y_k  + \tfrac{1}{2}\Lambda\epsilon_{ij}n_jn_kY_k\big)\,,\\
D_{Y_iY_j}&=&\ku^2\Big[\tfrac{3}{2}\big(2- 2\Lambda + \Lambda^2+Y_kY_k\big)\delta_{ij} - \big(1+2\Lambda-2\Lambda^2\big)Y_iY_j\,.
\\
&&\nonumber
+2\Lambda n_kY_k\big(n_iY_j+n_jY_i\big)
+6\Lambda n_in_j\Big]\,.
\end{eqnarray}
\end{subequations}
In the corresponding diffusion equation, we
change to polar coordinates $\ve Y\equiv Y[\cos\alpha,\sin\alpha]^\T$.
Since the flow is isotropic, we can remove one angular degree of freedom, leaving a two-dimensional diffusion equation in terms of $Y$ and $\delta\alpha\equiv\alpha-\psi$. We write the steady-state solution as $P(Y,\delta\alpha) = g(Y,\delta\alpha) Y$. In the limit of large $\Lambda$
the steady-state equation for $g$ takes the form:
\begin{align}
0&=24Y^2g
+3Y(1 + 8Y^2)\partial_Yg
+Y^2(3 + 4Y^2)\partial^2_Yg
+(3 + Y^2)\partial^2_{\delta\alpha} g\,
\end{align}
To solve Eq.~\eqnref{eq:diff_f} we attempt a Fourier expansion
\begin{equation}
g(Y,\delta\alpha)=\sum_m g_m(Y)e^{2\pi i m\,\delta\alpha}\,.
\end{equation}
This ansatz results in:
\begin{align}
0&=
[24Y^2-4\pi^2 m^2(3 + Y^2)]g_m
+3Y(1 + 8Y^2)\partial_Yg_m
+Y^2(3 + 4Y^2)\partial^2_Yg_m\,.
\eqnlab{diff_g_m}
\end{align}
The solution of this equation has power-law tails for large $Y$. The large-$Y$ asymptote of the solution can be written as a combination of two independent power-laws
\begin{equation}
\label{eq:gm}
g_m\sim c_mY^{-5/2 + 1/2 \sqrt{1 + 4 m^2 \pi^2}}+d_mY^{-5/2 - 1/2 \sqrt{1 + 4 m^2 \pi^2}}\,.
\end{equation}
This solution gives a normalisable probability distribution $P_m=g_mY$
for large values of $Y$ if $m=0$.
However, solutions with $c_m=0$ and $m\ne 0$ diverge as $g_m\sim Y^{-2\pi m}$ for small $Y$, leaving only $m=0$ as a valid solution. It follows that
the leading large-$Y$ asymptote of $g_0$ is $g_0(Y) \sim Y^{-3}$.
The general $m=0$-solution of \Eqnref{eq:diff_g_m} is
\begin{equation}
g_0(Y) =
c_0\frac{\sqrt{1+\tfrac{4}{3}Y^2}-\mbox{acoth}\big(\sqrt{1+\tfrac{4}{3}Y^2}\big)}{[1+\tfrac{4}{3}Y^2]^{3/2}}
+ \frac{4}{3}\frac{d_0}{[1+\tfrac{4}{3}Y^2]^{3/2}}\,.
\end{equation}
Using $c_0=0$ and $d_0=1$ the normalized probability distribution of angle gradients takes the form
\begin{equation}
\label{eq:supp_pY}
P(Y) = \frac{\tfrac{4}{3}Y}{[1+\tfrac{4}{3} Y^2]^{3/2}}\,,
\end{equation}
for large $\Lambda$.
The distribution of $Y_i$ is obtained
by noting that the joint distribution $P(Y,\delta\alpha,\delta\psi)$
is uniform in $\delta\alpha$ and $\delta\psi$. Therefore the distribution
$P(Y,\alpha)$ is uniform in $\alpha$. It follows that
$P(Y_1,Y_2)=P(Y)/(2\pi Y)$ with $Y=\sqrt{Y_1^2+Y_2^2}$.  Integrating over $Y_2$ gives
\begin{equation}
P(Y_1)=\frac{2}{\sqrt{3}\pi(1 + \tfrac{4}{3} Y_1^2)}
\end{equation}
This distribution has the $P(Y_1) \sim Y_1^{-2}$ tails mentioned in the main text.

\section{Calculation of the Lyapunov dimension $D_{\rm L}$}
\label{app:DL}
The Lyapunov dimension is a measure of the fractal dimension of the attractor in phase space.
In two spatial dimensions, phase space is three-dimensional ($x_1$, $x_2$, and $\psi$), so that there are three Lyapunov exponents, $\sigma_1\ge\sigma_2\ge\sigma_3$.
The sign of the maximal Lyapunov exponent $\sigma_1$ determines whether small separations grow (positive sign) or shrink (negative sign) exponentially.
The signs of partial sums of the $n$ upper Lyapunov exponents $\sigma_1+\sigma_2+\dots+\sigma_n$ determine whether $n$-dimensional sub-volumes of phase space grow or shrink exponentially.
Since the underlying flow is incompressible, the fractal dimension cannot be smaller than two (where $\sigma_1+\sigma_2=0$ and $\sigma_3<0$),
 and it cannot exceed three, the dimensionality of phase space
(where $\sigma_1+\sigma_2+\sigma_3=0$).
The Lyapunov dimension is defined as the linear interpolation between these limits \cite{Som93,Gus16a}:
\begin{align}
D_{\rm L} \equiv 3-\frac{\sigma_1+\sigma_2+\sigma_3}{\sigma_3}\,.
\eqnlab{supp_DL_def}
\end{align}
To evaluate the Lyapunov exponents in \Eqnref{eq:supp_DL_def}, we first note that two phase-space Lyapunov dimensions are given by the spatial Lyapunov exponents $\sigma_1^{\rm spatial}$
because the spatial dynamics is not influenced by the angular dynamics.
Also $\sigma_2^{\rm spatial}=-\sigma_1^{\rm spatial}$.
This equality follows from incompressibility of the flow.

As mentioned above, the Lyapunov exponents are ordered with respect to their size. We discuss this ordering below. For the moment we refer to the remaining
exponent as $\sigma'$.  It is determined from the local dissipation in phase space (and made dimensionless using $\tau$):
\begin{align}
\sigma' =  \sigma_1^{\rm spatial}+\sigma_2^{\rm spatial}+\sigma' =
\sigma_1+\sigma_2+\sigma_3&=\langle\partial_x\dot x+\partial_y\dot y+\partial_\psi\dot\psi\rangle=-2\ku\Lambda\langle\ve n\cdot\ma S\ve n\rangle\,.
\end{align}
Using perturbation theory
for small values of $\ku$, the spatial Lyapunov exponents $\sigma_\mu^{\rm spatial}$
were obtained in Ref.~\cite{Gus16a}
(Eq.~(114) in that paper, evaluated for $\st=0$, $d=2$, and $\mu=1,2$). To order $\ku^4$ one finds:
\begin{align}
\sigma_1^{\rm spatial}\equiv\ku\langle\hat{\ve R}\cdot \ma S\hat{\ve R}\rangle  = \ku^2 -6 \ku^4\,,\quad\mbox{and}\quad
\sigma_2^{\rm spatial} = -\sigma_1^{\rm spatial}\,.
\end{align}
where $\hat {\ve R}$ is the unit separation vector between a pair of particles.
Using a similar expansion one can also calculate  $\sigma'$:
\begin{align}
\sigma'\equiv -2\ku\Lambda\langle\ve n\cdot\ma S\ve n\rangle = -2\ku^2\Lambda^2+4\ku^4\Lambda^2 (2+\Lambda^2)\,.
\end{align}
We remark that for $\Lambda=1$, both $\hat{\ve R}$ and $\ve n$ follow the same dynamics. This implies $\sigma'=-2\sigma_1^{\rm spatial}$ for $\Lambda=1$.
Before inserting the exponents into \Eqnref{eq:supp_DL_def}, we must order them.
We find that $\sigma_1=\sigma_1^{\rm spatial}$ for any value of $\Lambda$ and that $\sigma_2=\sigma'$ and $\sigma_3=\sigma_2^{\rm spatial}$
if $\Lambda<\Lambda_{\rm c}$, where
\begin{equation}
\Lambda_{\rm c}=\tfrac{1}{\sqrt{2}}(1 -\tfrac{1}{2} \ku^2)\,.
\end{equation}
 If $\Lambda>\Lambda_{\rm c}$ we instead have $\sigma_2=\sigma_2^{\rm spatial}$ and $\sigma_3=\sigma'$.
Inserting these ordered Lyapunov exponents into \Eqnref{eq:supp_DL_def}, we obtain to second order in $\ku$
\begin{equation}
D_{\rm L} =
\left\{
\begin{array}{ll}
3- 2\Lambda^2 + 4\ku^2\Lambda^2(\Lambda^2-1) & \mbox{ for } \Lambda<\Lambda_{\rm c}\,,\cr
2 & \mbox{ for }\Lambda\ge \Lambda_{\rm c}\,.\cr
\end{array}
\right.
\label{eq:supp_DL}
\end{equation}
This is Eq.~(\ref{eq:result_DL}).

\end{document}